# A Comprehensive Approach to Abusing Locality in Shared Web Hosting Servers

Seyed Ali Mirheidari[1], Sajjad Arshad[2], Saeidreza Khoshkdahan[3], Rasool Jalili[4]

[1]Computer Engineering Department, Sharif University of Technology, International Campus, Kish Island, Iran
[2]DNSL, Computer Engineering Department, Sharif University of Technology, Tehran, Iran
[3]Sabzfaam Information Technology Corporation, Tehran, Iran
[4]Computer Engineering Department, Sharif University of Technology, Tehran, Iran

mirheidari@kish.sharif.edu, msarshadir@gmail.com, khoshkdahan@sabzfaam.ir, jalili@sharif.edu

*Abstract*—**with the growing of network technology along with the need of human for social interaction, using websites nowadays becomes critically important which leads in the increasing number of websites and servers. One popular solution for managing these large numbers of websites is using shared web hosting servers in order to decrease the overall cost of server maintenance. Despite affordability, this solution is insecure and risky according to high amount of reported defaces and attacks during recent years. In this paper, we introduce top ten most common attacks in shared web hosting servers which can occur because of the nature and bad configuration in these servers. Moreover, we present several simple scenarios that are capable of penetrating these kinds of servers even with the existence of several securing mechanisms. Finally, we provide a comprehensive secure configuration for confronting these attacks.**

## I. INTRODUCTION

Some companies and organizations use dedicated web servers while the progression of hardware enables multitudes of websites to reside on one server. This solution is commonly known as Shared Web Hosting which has several advantages including affordability and using maximum power of server hardware. Another advantage of this solution is that it is not required for the website owners to be experts of the domain and they can only handle their own website applications.

The risks and vulnerabilities involved with using shared web hosting can prevent it from always being an excellent solution. As claimed by Zone-H, an unexpected number of successful attacks were fulfilled after accessing a vulnerable website on a shared web hosting server and even secure websites like static pages also being hacked due to residing on shared web hosting servers [1] [2]. Various security challenges in different levels of network come along with these kinds of servers because there is no proper isolation between resources used by websites [3]. Also, having one vulnerable website on the shared server allows the attacker to hack into other websites with no struggle due to improper configuration of shared web hosting servers.

In this paper, we present top ten most common attacks in shared web hosting installations and then provide a comprehensive secure configuration for shared web hosting servers. A specific configuration is needed for the shared hosting servers to become vulnerable to these attacks [3]. In this way, an attacker who controls a website hosted on a shared web hosting server is able to attack all other websites hosted on the same server.

In this paper, we focus on the Apache webserver to present the attacks. Apache webserver as mentioned in Netcraft [4], is the most common used webserver among other webservers such as Microsoft IIS. In addition, the focus of this paper is on Linux operating system due to the fact that most countermeasures are developed for the POSIX operating system. Also, we use PHP programming language because of higher popularity, usability and reliability. However, it has to be known that any webserver with certain configuration can be vulnerable to the aforementioned attacks and they are not only for the Linux/Apache/PHP installation.

In summary, this paper provides the following contributions:
- We demonstrate top ten attacks in shared web hosting servers where six of these attacks are novel.
- We provide sample codes to clarify the concept of these attacks.
- We provide a comprehensive configuration for shared web hosting servers to encounter these attacks.

The remainder of this paper is outlined as follows. In Section II, the overall architecture of shared web hosting servers is portrayed. We describe top ten most common attacks against shared web hosting servers in Section III. In Section IV, we present a comprehensive secure configuration for shared web hosting servers to defeat these attacks and we conclude in Section V.

## II. SHARED WEB HOSTING ARCHITECTURE

In order to gain a more precise view for understanding the attacks mentioned in the next section, we describe the shared web hosting architecture in this section. In shared web hosting, a webserver hosts many websites simultaneously. There is a FTP account for every website owner for uploading new files which are owned by the user account of the website owner. Webserver runs as a specific user account (apache, daemon, www-data) and handles all HTTP requests for all websites. Thus, it is necessary for a webserver to have the ability to read the files of every website. Despite this, users in some Content Management Systems (CMS) must be capable of uploading files thus besides reading files, write access to files and directories of websites is also needed for a webserver [5].

In Figure 1 where web1 and web2 are owners of two separate websites, necessary permissions are shown for the Apache webserver in Linux operating system.

```
d rwx r-x ---    web1:www-data    /home/website1/public_html
d rwx r-x ---    web2:www-data    /home/website2/public_html
```

Figure 1. Essential Permissions for the Apache Webserver

There are two universal forms of webserver configurations for executing scripts in shared web hosting, which are:
- Configuring the webserver to load the script interpreter as a webserver module
- Configuring the webserver for running the script interpreter as a CGI [6] binary

The webserver process loads the webserver module or it is compiled into webserver binary, meaning that a binary image of the interpreter exists in the webserver process. In CGI mode, by arrival of each request, the webserver will create a new interpreter process to handle it. In comparison to the CGI mode, using script interpreter as webserver module has advantages like more stability under load and more efficiency in managing requests and resources. However due to the fact that malicious scripts do not affect webserver process, the CGI mode is more secure.

### III. SHARED WEB HOSTING ATTACKS

In Section II, we provided an overview of how shared web hosting servers work. The attacks which we are going to present are based on the fact that there is no proper isolation between different websites hosted on the shared web hosting server.

This weakness gives the attacker a set of capabilities like accessing files of other websites and exploiting the disclosed information (Data Confidentiality Violation), modify important files (Data Integrity Violation), force arbitrary sessions to the vulnerable websites (Session Poisoning), inspect and modify their session values (Session Snooping), manipulate logs of other websites (Log Poisoning), inspect their logs (Log Snooping), execute malicious code by using LFI (Intensive LFI), inspect and modify CSRF Tokens (CSRF Token Poisoning), launch brute force attack faster (Fast Brute Force) and easily phish victims (Convenient Phishing).The details of these attacks are presented in following sections.

#### A. Data Confidentiality Violation

In shared web hosting servers, webserver is run by one individual user account in a default way and the scripts of all websites are executed under that user account. Thus, this user account can access all files and folders and as a result all scripts of a website can access the files and folders of other websites. One attacker can access the files and folders of a website belonging to the victim and exploit the resulted information [5].

For instance, an attacker can generate malicious script and read the database configuration file which includes username and password (which are usually in clear text) and connect to database and read private data. Using this method, an attacker can exploit the resulted information and change the behavior of the website in a desired way.

#### B. Data Integrity Violation

As mentioned before, in shared web hosting servers users have read access to all files of all websites. But there are some websites like CMS that give the user the ability of uploading files. In this case if database is not used for storing uploaded files, the webserver user account should have permission for creating new files in folders of one individual website. In other words, webserver user account has read/write access. Since in the default mode, scripts of all websites run under webserver user account, by using this method attacker can recognize victim websites and change it in a desired way [5].

As an illustration, an attacker can write a script to search files with the read permission and find the vulnerabilities of websites and finally attacker can use another script to modify important files or create some new files in victim websites.

#### C. Session Poisoning

In shared web hosting servers, websites store their session files in temp directory based on their needs and all users can read or modify the files in temp directory. Thus, there is a strong possibility for users to have access to session files of a website in the case we do not use secure mechanisms.

In Session Poisoning [7] attack, attackers create a new session file and force the victim site to use it like other valid session files. As an illustration assume that the victim website has an administration panel with an authentication page. When administrator login the page successfully, victim website defines a new session variable named as *level* and send session ID to the admin client. Thus, in future it is enough for administrator to send just the session ID and not all the login information. On the other side, an attacker can create a session file using his own website which includes the admin *level* variable. Finally, the attacker will see the administration panel of the victim website which sends the newly created session ID to the victim website. Since both websites hold their session files in temp directory, the victim website loads the session file uploaded by the attacker and consequently gives the attacker administration permission. This can be also done by the attacker with inserting one fake session in the temp folder if it is possible for him to write directly on the files of temp folder.

#### D. Session Snooping

In shared web hosting servers with default configuration, all websites store their sessions in one directory (e.g. */tmp*). In other words, all websites can read or modify session files.

In Session Snooping [7] attack, attackers analyze and modify the content of the session files of other website in order to exploit the results. For example, assume one forum as a victim website in which users can login and access different pages based on their permissions. The victim website uses *username* variable in its session in order to prepare appropriate results for different users. On the other side, an attacker registers itself in victim forum and login successfully afterwards which leads in sending session ID from victim website to attacker. Then the attacker changes the value of correspondent *username* variable in new created session file to the username of the victim. Finally, the attacker reviews the victim websites and sends his session ID to it. Consequently, victim website loads the session file related to the session ID and recognizes the attacker as another user. Thus, the attacker can visit private information of other users.

### E. Log Poisoning

Webservers usually save the information of processed requests in a log file. A log file includes information such as Domain Name, Client IP, Request Time, Request Type (GET or POST), Requested Filename, Size of Transferred File and Return Status Code from webserver [8]. The two attacks presenting are based on the fact that Webserver uses a single file for storing logs of various websites and the log file is accessible by every script executed by the webserver [9]. In the following sections, the details of Log Poisoning and Log Snooping attacks will be presented.

In default configuration of a shared web hosting server, modifying the log file is only allowed by the root user and other users can only read it. Also, permission is required for a webserver to write in the log file regardless of the user account running with it. Therefore, in most webservers like Apache, parent webserver is executed with root privilege and child webservers are run by parent webserver to handle the requests. In Linux and other Linux-like operating systems, child processes inherit file descriptors opened by their parent process. Now by opening a file by a parent process in write mode, child processes are able to write in the already opened file. In this way, although not having root privilege, child webservers inherit the log file descriptor and can alter the log file. Scripts of websites are able to modify the log file since they are executed by child webservers.

In Log Poisoning [9], a script is created by the attacker in order to find log file descriptor and open the log file in write mode. In Linux operating system for example, information about opened files of each process exists in */proc/PID/fd*, where PID is the process ID. A PHP script is then created by an attacker for finding opened files of child webserver processes and the script is executed and the log file is opened again but with write access. The sample PHP script for Log Poisoning attack is displayed in Figure 2.

In order to be susceptible to this attack, PHP interpreter must be used as an Apache module in Apache webserver because log file descriptor is not inherited by the new CGI interpreter process and as a result the log file cannot be re-opened in write mode by the malicious PHP script. Clearing other websites' requests for covering track of penetration and adding fake requests to the log file are examples of malicious activities which an attacker can carry out in case of having write access to the log file. We must emphasize that from a general view having write access to log file in shared web hosting will lead to very dangerous situation which attackers can fulfill various attacks on the hosted websites.

```php
<?php
  if ($dh = opendir('/proc/self/fd/')) {
    while (($fd = readdir($dh)) !== false) {
      if (strpos(realpath($dir.$fd), "access_log") !== false) {
        $log_fd = $fd;
        break;
      }
    }
    closedir($dh);
  }
  $file = fopen("php://fd/$log_fd", "w");
  fwrite($file, "Some Junk Data\n");
  ...
  fclose($file);
?>
```

Figure 2. Log Poisoning Attack Script (PHP-Module Mode)

### F. Log Snooping

In default configuration, all scripts run by the webserver can read the log file because webserver user account has read access to log file. Therefore, scripts of one website are capable of reading logs of other websites located on the same shared web hosting server. In Log Snooping [9] attack, the goal of an attacker is to retrieve critical information by searching the other website' logs in order to launch other complex attacks. Unlike Log Poisoning, Log Snooping attack is feasible in two modes which webserver runs the script interpreter (Module or CGI).

```php
<?php
  if ($dh = opendir('/proc/self/fd/')) {
    while (($fd = readdir($dh)) !== false) {
      if (strpos(realpath($dir.$fd), "access_log") !== false) {
        $log_fd = $fd;
        break;
      }
    }
    closedir($dh);
  }
  $file = fopen("php://fd/$log_fd", "r");
  $data = fgets($file);
  ...
  fclose($file);
?>
```

Figure 3. Log Snooping Attack Script (PHP-Module Mode)

```php
<?php
  $file = fopen("/var/log/apache/access_log", "r");
  $data = fgets($file);
  ...
  fclose($file);
?>
```

Figure 4. Log Snooping Attack Script (PHP-CGI Mode)

Even when configuring the log file as unreadable for other users, Log Snooping attack can be done using the PHP script shown in Figure 3 and Figure 4. Structure of files and folders of victim websites is one of the most important information that attackers can acquire by Log Snooping attack. Attackers can re-generate the site tree using requested URLs and find out about names of website files and folders.

In several hardening best practices, the name of administrator authentication page is changed to prevent attackers from penetrating in the administration panel. Attackers can bypass this technique and find the authentication page by using the site tree. So, the attacker can use methods like SQL Injection to extract hashed password of administrator and find clear password text by using brute force of encoded password or using brute force for both user and password to discover the admin login credentials. A significant fact here is that the attacker cannot easily find the authentication page in case of not having access to the shared web hosting server.

*G. Intensive Local File Inclusion (LFI)*

Some websites are vulnerable since they allow special code reuse by including files through supplying the values of some parameters in URL. In this case, attackers try to misuse and include some malicious files. One of the most common attacks in this area is known as Local File Inclusion (LFI) [10] which leads in including victim website's local file. During recent years, several methods like LFI2RCE [11] have been proposed which are able to execute remote code using LFI vulnerability. One method is to inject malicious code into the log file of webserver and include the log file by LFI which leads to execution of malicious code by victim website.

But in case of large log file, this method is not effective due to the fact that websites cannot include the whole file. Generally, we can say without having access to the local victim file system, it is a complicated task to execute malicious code by including common files such as log file. But in shared web hosting servers, it is easier since there is an access to local victim file system and an attacker can do the LFI2RCE attack easily. As an illustration, attackers can create a malicious file in a path like temp directory which can be accessed by all websites and use the LFI attack to include malicious code and consequently execute its malicious code. Thus, we can conclude that in shared web hosting servers, LFI2RCE attack is more common than its alternatives.

*H. Cross Site Request Forgery (CSRF) Token Poisoning*

Cross Site Request Forgery (CSRF/XSRF) [12] is a type of vulnerability in a website whereby unauthorized commands are transmitted from a user trusted by the website. In other words, CSRF exploits the trust a website has in the browser of users [13]. An easy and effective solution is to use a secret, user-specific and server-side generated token [12]. In this way, websites generate a token and sends it to the browser of the user. After that, the browser should use the token in all form submissions. When the website receives a request from a user, it checks the received token with the original one and if two tokens match, it will process the request of the user. Whereas attackers are not able to put the right token in their submissions, they cannot launch CSRF attack.

Websites usually generate tokens per-session and save the token values in the corresponding session files. Attackers create some scripts to inspect and modify content of session files belonging to other websites in shared web hosting servers. As a result, attackers are able to modify the value of CSRF tokens which are located in session files. In this way, attackers can bypass CSRF prevention technique and send unauthorized requests to the victim websites.

*I. Fast Brute Force*

Nowadays, Brute Force [14] attack is known as a common attack on web applications for detecting passwords, directories, files and session IDs. Bandwidth protection and request controller tools are the main constraints of this type of attacks. In dedicated servers, attackers must follow remote attack in order to detect the password and most of attempts fail because of the low bandwidth.

But in shared web hosting servers, attackers can try brute force attack locally without any bandwidth limit. Moreover, since shared servers usually have high processing power, attackers can use the CPU along with their goals and create a script including brute force code and start the attack on the victim website. So in shared web hosting servers, brute force attacks tried locally are generally faster than remote ones on dedicated servers. Trying local attempts not only bypass the bandwidth bottleneck but also bypass the protection mechanisms.

*J. Convenient Phishing*

Phishing is a kind of online identity theft in which confidential information of users such as bank account password or credit card information is stolen by displaying fraudulent web pages. Nowadays Phishing is an important attack on the internet and is accepted as a global criminal activity. In simple words, phishers try to redirect the users to a website where they are asked to enter the personal information. E-mail and online banking websites are the main target of phishing. Fake websites are designed in a way to seem as a legitimate website and afterwards phishers use the private information for malicious tasks [15].

Currently there are some techniques for protecting users against phishing [16]. But in shared web hosting servers, phishers can bypass many phishing prevention mechanisms because they can access to webservers of victims. For instance, users may use address *www.website.com/~attcker/page.php* for login instead of *www.website.com/~victim/page.php*. Interestingly, the prevention mechanisms cannot detect the anomaly activity since the domain address is the main domain of website and homographic domains have not been used. In other words, with access to shared web hosting servers, attackers are able to easily bypass many phishing obstacles.

IV. A COMPREHENSIVE SECURE CONFIGURATION

In this section, we present a secure configuration for Linux/Apache/PHP installations and how it confronts the attacks described in Section III. Precisely observing and studying these attacks leads us to the fact that the rise of such attacks is due to the lack of proper isolation between different websites hosted on one server [17]. Following sections provide details of proposed configuration for shared web hosting servers.

*A. Script Execution Restriction*

In default shared web hosting configuration, all scripts are executed under the user account of the webserver regardless of their owners. Thus, a website is able to access resources of other websites. Due to the popularity of shared web hosting, several methods have been presented for providing a more secure shared web hosting installation. Following sections introduce the most well-known countermeasures developed for this purpose.

*1) PHP Methods*

Safe_Mode [18] and Open_Basedir [19] are two methods which PHP developers are studying in order to solve the security problem, although they both carry on some limitations. In other words, PHP is not the right platform for unraveling the security problem [5].

**Safe_Mode.** In Safe_Mode, PHP examines the access of running PHP scripts to files based on their owners. PHP checks the owner of those files and if the owner of the file is not the same as the owner of the running script, PHP will not allow that access. However, Safe_Mode has a few limitations. It has to be known that some applications that upload files to server, the owner of them will be Apache user, not the script owner's user account and those files cannot be accessed by the PHP scripts anymore [5].

**Open_Basedir.** In Open_Basedir, PHP determines the directory which each user is allowed to access. PHP examines the file access of running PHP scripts and do not allow access to files outside that directory [5].

*2) Apache Module Methods*

By precisely seeing the security problem, the cause of the problem will rise as how we run the Apache server. Apache is executed by a unique user who can have access to all files of all websites.

A new idea is that Apache can serve each website by its owner's user account. In other words, each script is run with its owner's user account permissions. suEXEC [20] and suPHP [21] are two well-known methods which use this idea and have been developed as an Apache module.

**suEXEC.** The suEXEC includes a wrapper binary file and an Apache module. By the arrival of a HTTP request, the wrapper is run by Apache and the script name and user/group ID under which the script has to be executed is given to the wrapper. The suEXEC can only be used with CGI or FastCGI programs. For using suEXEC, a unique CGI or FastCGI binary file for each website is needed. The user/group ID of the owner must be the website's owner. By the release of a new PHP version, these binary files must be updated and in case of using PHP in CGI or FastCGI mode, HTTP authentication feature cannot be used. Using suEXEC with CGI has very low performance in a way that Corentin Chary has named suEXEC a performance killer due to tis low performance in use with CGI [5] [22].

**suPHP.** Same as suEXEC, suPHP runs PHP scripts with a specified user/group ID. The suPHP has an Apache module and a setuid-root binary file. Unlike suEXEC, using suPHP will not require a unique CGI or FastCGI [23] binary file for each website. The low performance issue still remains in suEXEC same as the suPHP [5] [24].

*3) Apache MPM Methods*

After the release of Apache 2.0, various MPM [25] methods have been introduced in order to solve the shared hosting security problem [5]. These methods are tested with greater details in the following sections.

**Peruser MPM.** Because Metux MPM [27] is not appropriate for PHP, Sean Gabriel Heacock introduced Peruser MPM [26]. Peruser MPM uses processes instead of threads to handle requests. Peruser MPM runs an Apache control process as root privilege. The control process creates several multiplexer processes with Apache user privilege. The multiplexer process listens to port 80 and accepts incoming connections and reads the request to check from which website it is and it passes the connection to related worker process to be handled. The worker processes run under the user/group ID of respective owners of websites. The control process always maintains a pool of idle worker processes to enhance the performance and forks off new worker processes if there are no idle processes to handle new requests. However, one important deficiency of Peruser MPM is too much use of server resources [5].

**ITK MPM.** Steinar Gunderson presented ITK MPM [28] to reduce the shortcomings of Peruser MPM. ITK MPM creates a managing Apache process with root privilege. The managing process creates several listener Apache processes with root privilege. The listener process listens to port 80 and reads new request to determine the requested website. In order to serve the request, the listener creates an Apache handler process with user/group ID of the owner of the website. But, the main difference between ITK MPM and Peruser MPM is that the handler Apache process is terminated after the request has been completed. In other words, in ITK MPM there is no pool of idle handler processes for serving the requests [5].

*B. Log File Separation*

Based on the fact that a webserver with default configuration will use a single file for logging activities of all websites, not having a proper separation between log files of different websites is the cause of log attacks [9]. The best solution for preventing these attacks is creating separate log file for each website and putting them in separate directories. Certainly proper permissions must be set on the log files to prevent a malicious user from reading or writing on log files of other websites [9]. A sample configuration in Apache webserver for creating separate log file for each virtual host or website is displayed in Figure 5. Also, the necessary permissions on log file directories in Linux are shown in Figure 6 where web1 and web2 are user accounts of owners of the corresponding websites.

*C. Session Storage Separation*

As same as log attacks, the main cause of session attacks is the lack of separation between session storage of different websites [7]. In case of Session Poisoning and Session Snooping attacks, a webserver with default configuration will use a temp directory for storing session files of all websites. Therefore, separating session storage for each website and providing separate directories for each one is necessary for stopping these attacks. In addition, proper permissions must be set on the session directories. Figure 5 shows sample of configuring an Apache webserver for creating separate session directories for each virtual host or website and in Figure 6 the required permissions are depicted.

```
<VirtualHost *:80>
  DocumentRoot   /home/website1/public_html
  ServerName     website1
  ErrorLog       /home/website1/log/error_log
  CustomLog      /home/website1/log/access_log common
  php_value      session.save_path /home/website1/session
  ...
</VirtualHost>

<VirtualHost *:80>
  DocumentRoot   /home/website2/public_html
  ServerName     website2
  ErrorLog       /home/website2/log/error_log
  CustomLog      /home/website2/log/access_log common
  php_value      session.save_path /home/website2/session
  ...
</VirtualHost>
```

Figure 5. Log and Session Separation in Apache for Each Website

```
d rwx r-x ---   web1:web1   /home/website1
d rwx r-x ---   web1:web1   /home/website1/public_html
d rwx r-x ---   web1:web1   /home/website1/log
d rwx r-x ---   web1:web1   /home/website1/session

d rwx r-x ---   web2:web2   /home/website2
d rwx r-x ---   web2:web2   /home/website2/public_html
d rwx r-x ---   web2:web2   /home/website2/log
d rwx r-x ---   web2:web2   /home/website2/session
```

Figure 6. Necessary Permissions for Log and Session Directories in Linux

### D. Local Access Limitation

In shared web hosting servers, the local host is usually trusted and consequently an attacker, who has a website on the shared server, is able to launch attacks such as Fast Brute Force. In order to prevent such attacks it is a good idea to control local traffic. In other words, the local traffic must be gone through security devices like WAFs and NIDPSes before reaching target website.

## V. CONCLUSION

Today, shared web hosting is recognized as a popular approach to host thousands of websites but it has multiple serious vulnerabilities which are primarily due to the fact that different resources such as memory, CPU, network and file system are shared between different websites.

In this paper we addressed common attacks in shared web hosting servers which exploit the lack of proper isolation between different websites resided on a shared server. Then, we presented a comprehensive secure configuration to prevent the risks of these attacks. As a conclusion we can say that although the mentioned mechanisms prevent the attacks directed towards the shared web hosting servers, but generally this architecture is not advised since it is potentially insecure and new solutions like virtualization are more secure and reliable.